# Sub-Wavelength Nonlinear Accelerating Beams

**Ido Kaminer and Mordechai Segev**
*Physics Department and Solid State Institute, Technion, Haifa 32000, Israel*
*msegev@tx.technion.ac.il*

**Abstract:** We show that optical nonlinearities allow sub-wavelength beams to propagate in circular trajectories *without being attenuated* in spite of their *partially evanescent spectrum*. Such beams are exact solutions to Maxwell's equations with Kerr or saturable nonlinearity.
**OCIS codes:** (070.7345) Wave propagation; (050.6624) Subwavelength structures; (260.2110) Electromagnetic optics; (350.7420) Waves; (190.0190); Nonlinear optics; (190.6135) Spatial solitons; (190.3270) Kerr effect

The research on accelerating beams has been growing rapidly since it was brought into the domain of optics in 2007 [1]. The best known accelerating beam is the paraxial Airy beam, which propagates along a parabolic trajectory while preserving its amplitude structure indefinitely. The effect is caused by interference: the waves emitted from all points on the Airy profile maintain a propagation-invariant structure, which shifts laterally along a parabola. This beautiful phenomenon has led to many intriguing ideas such as guiding particles along a curve [2], and to recent studies on shape-preserving accelerating beams in nonlinear optics [3-7]. The idea of nonlinear accelerating beams goes back to Giannini and Joseph [8] who found temporal self-accelerating solitons in Kerr media. Early attempts to launch Airy beams into nonlinear media have shown that the beams do not propagate as an accelerating self-trapped entity [3]. However, as we have shown recently [4], a specific design of the beam profile can make it accelerate in a shape-preserving manner while propagating in the nonlinear Kerr [4,5] nonlocal [6], and quadratic [4,6] media.

Until recently, in all studies on this subject, the accelerating beams were solutions of the (linear or nonlinear) paraxial wave equation. Therefore, the propagation had to stay paraxial, i.e., the trajectory was fundamentally limited to small (paraxial) angles. This restriction is a serious limitation, because the paraxial accelerating beam is moving on a curve which bends ever faster, and is eventually bound to break its own domain of existence. A recent study [8] used caustics to design beams that could bend to large non-paraxial angles, but such method inherently cannot really find shape-preserving solutions, hence the non-paraxial accelerating beams in [8] were not propagation-invariant. Finally, our recent work [9] presented the first non-paraxial shape-preserving accelerating beams that are solutions to the full Maxwell equations. Among other things, we have found that non-paraxial accelerating beams can be designed to have features on the scale of a single wavelength. This stretches the spatial spectrum of the beam to the maximum possible, without crossing the line to the evanescent regime. But even so, the fundamental limit remains: the features of the beams cannot be smaller than a single wavelength without being composed of evanescent waves. Otherwise, the evanescent part would inevitably decay, causing the beam to change its shape dramatically. This raise an intriguing question: are we restricted by the same limitation in **nonlinear** optics? Is it possible to design a **nonlinear sub-wavelength accelerating beam** which would not decay?

Here we present the first nonlinear self-trapped accelerating beams of the full (time harmonic) Maxwell equations. These beams accelerate along a circular trajectory in a shape-preserving manner, without decay or reflections. This accelerating shape-preserving propagation occurs even when the beam is constructed of sub-wavelength features, such that a significant part of its power resides in the evanescent regime. In this context, this work proves that the nonlinearity takes an exclusive role in the propagation, preventing the evanescent spectrum from decaying. Finally, we show that such nonlinear non-paraxial accelerating beams involve coupling between forward- and backward-propagating beams, giving rise to unique effects related to nonlinear evanescent waves.

We begin from Maxwell's equations with nonlinearity of the form $n^2(E)=n_0^2(1+\varepsilon(|E|))$ for a general saturable or Kerr-like nonlinearity that depends only on the intensity $I=|E|^2$. For a TE-polarized time-harmonic field $\vec{E}=E(x,z) \exp(ik_0ct) \hat{y}$ we get the nonlinear Helmholtz equation:

$$E_{xx} + E_{zz} + k_0^2 E + k_0^2 \varepsilon(I)E = 0. \qquad (1)$$

Equation 1 displays full symmetry between the x and z coordinates. Hence it is logical to seek a shape-preserving beam whose trajectory resides on a circle. We therefore transform to polar coordinates $r,\theta$ by taking $z = r\sin(\theta)$, $x = r\cos(\theta)$, and seek shape-preserving solutions of the form $E = U(r)e^{i\alpha\theta}$. Here, $\alpha$ is some real number indicating the phase velocity of the beam in the azimuthal direction, which is the desired direction of propagation. The radial function $U(r)$ must satisfy:

$$\frac{d^2}{dr^2}U + \frac{1}{r}\frac{d}{dr}U - \frac{\alpha^2}{r^2}U + k_0^2 U + k_0^2 \varepsilon(U)U \,. \qquad (2)$$

A special case of this equation (specific nonlinearity and no acceleration; $\alpha=0$) is solved in [10], giving the nonlinear Bessel beam with zero angular momentum. In contrast to that, we are interested in high values of $\alpha$ (large beam bending). To find accelerating ($\alpha>0$) solutions of Eq. (2), we notice that for some sufficiently small $r$ value (typically $\alpha/k_0$), the amplitude is small enough such that the nonlinear term is negligible. Hence, we use the analytic

Bessel solution of the linear equation to set the boundary conditions: $U(r_0)=J_\alpha(r_0 k_0)$, $U'(r_0)=J'_\alpha(r_0 k_0)$. This method successfully finds solutions for the Kerr nonlinearity in both focusing and de-focusing cases, and also works for saturable nonlinearities. Other types of nonlinearity would also yield solutions to, but certain cases of strong nonlinearities will have a threshold for the existence of an accelerating solution. For example, above a certain value of defocusing Kerr, no solution exists (a similar effect is found in [4]). Figure I below shows the shape of a nonlinear accelerating beam, compared with the linear accelerating beams; both are cut at the same finite aperture.

In all cases, the result is a beam which is shape-preserving along any circular curve (Fig.II). One might be tempted to assume this beam propagates in circles endlessly. However, the boundary conditions are still in $x,z$: the incident beam is launched as $E(x,z=0)$ and is propagating in the forward $+z$ direction. Therefore, the solution of (2) should be used only as the initial condition (launch beam) for $x>0$ in the $z=0$ plane. This selection of initial condition breaks the full symmetry of the polar coordinates, and limits the acceleration of the beam to one quarter of the circle, after which the trajectory is bending back toward $-z$. Only the backscattered wave is allowed to go back along the second quarter of the circle. Consequently, the exact solution of Eq.(2) is not the actual physical solution for a beam launched at $z=0$, but instead, it approximates the desired beam with increasing accuracy for larger $\alpha$. This peculiarity is discussed in more details in [9], although an important difference should be noted: in linear media, the forward and backward propagating waves are completely decoupled, whereas in our nonlinear case here the forward and backward propagating waves interact through the nonlinearity, hence they cannot be separated.

Figure III presents the forward propagating beam, calculated by the projection $-\frac{i}{2}\sqrt{k_0^2-k_x^2}\partial_z+\frac{1}{2}$ from the exact solution of (2) into its forward propagating wave (compare to the full wave in Fig.II). This projection is taken in the Fourier plane ($x \rightarrow k_x$) and is exact only when the nonlinearity is small.

Figure IV displays the Fourier transform of the beams of Fig.I. The dashed line indicates the edge of the evanescent regime, emphasizing that the spatial spectrum of the nonlinear beam is largely evanescent, and thus it relies on the nonlinearity to propagate without decaying. This finding conforms with the thin sub-wavelength lobes shown in Fig.I, where even the main lobe is about half a wavelength wide. As a consequence, when the beam completes near 90° bending, it breaks up, causing the nonlinearity effect to drop quickly, causing most of the beam to decay. For stronger nonlinearities, we expect an even more abrupt decay, causing a larger portion of the power to backscatter. This breaks another fundamental rule of linear optics: linear propagation prohibits the backward radiation to be created from an incident beam in a homogenous medium; and yet, nonlinear optics allows the nonlinear beam to reflect itself (or part of it) – which is expressed in an effective bend of **more than 90°**. This beautiful phenomenon is at its best in the nonlinear dynamics of our accelerating beams, since the reflection changes from zero, along the propagation, to its maximum, right after the nonlinearity drops.

[1] G. A. Siviloglou and D. N. Christodoulides, Opt. Lett. **32**, 979 (2007); Phys. Rev. Lett. **99**, 213901 (2007).
[2] J. Baumgartl, M. Mazilu and K. Dholakia, Nature Photonics **2**, 675 (2008).
[3] Y. Hu, S. Huang, P. Zhang, C. Lou, J. Xu, and Z. Chen, Opt. Lett. **35**, 3952 (2010).
[4] I. Kaminer, M. Segev, and D. N. Christodoulides, Phys. Rev. Lett. **106**, 213903 (2011).
[5] A. Lotti, D. Faccio, A. Couairon, D. G. Papazoglou, P. Panagiotopoulos, D. Abdollahpour and S. Tzortzakis, Phys. Rev. A **84**, 021807 (2011).
[6] R. Bekenstein and M. Segev, Opt. Express **19**, 23706 (2011).
[7] I. Dolev, I. Kaminer, A. Shapira, M. Segev, and A. Arie, accepted to publication in Phys. Rev. Lett. (2012).
[8] L. Froehly, F. Courvoisier, A. Mathis, M. Jacquot, L. Furfaro, R. Giust, P. A. Lacourt, and J. M. Dudley, Opt. Express **19**, 16455 (2011).
[9] I. Kaminer, R. Bekenstein, J. Nemirovsky and M. Segev, arXiv:1201.0300v2 (2012).
[10] M. A. Porras, A. Parola, D. Faccio, A. Dubietis, and P. Di Trapani, Phys. Rev. Lett. **93**, 153902 (2004).

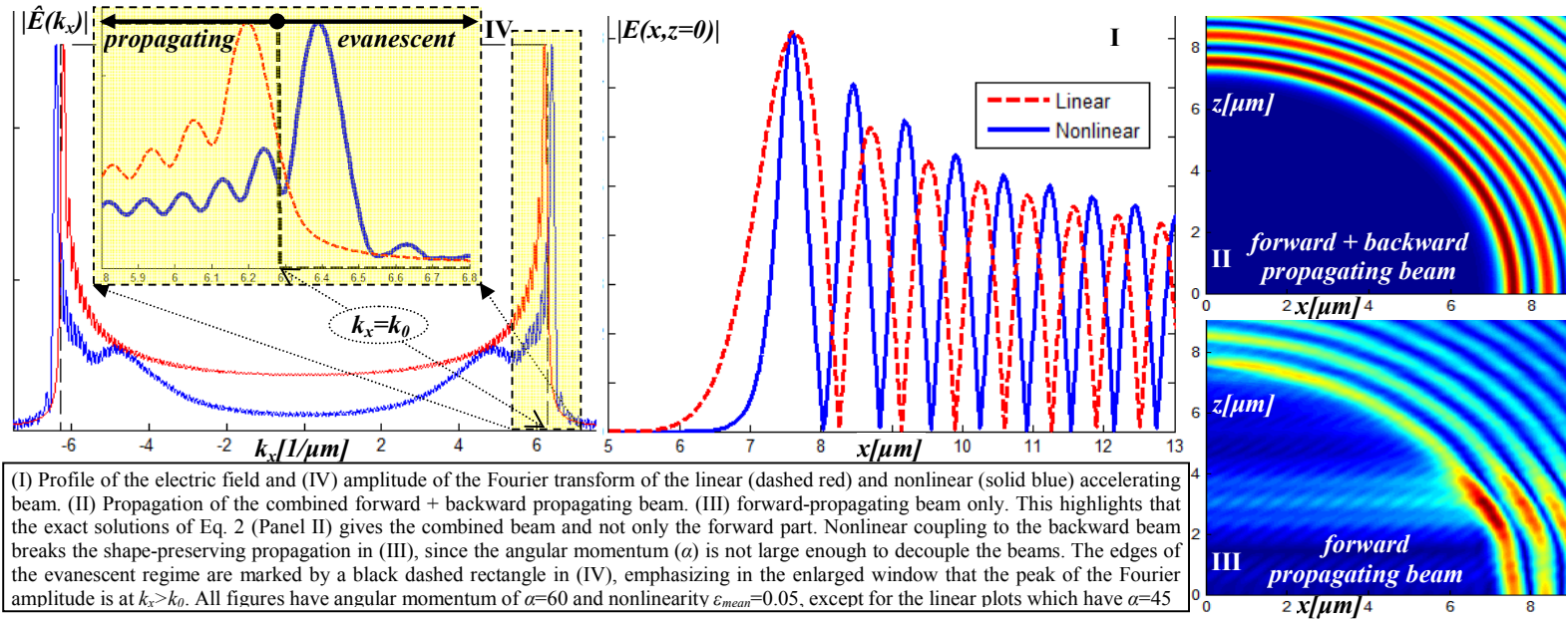

(I) Profile of the electric field and (IV) amplitude of the Fourier transform of the linear (dashed red) and nonlinear (solid blue) accelerating beam. (II) Propagation of the combined forward + backward propagating beam. (III) forward-propagating beam only. This highlights that the exact solutions of Eq. 2 (Panel II) gives the combined beam and not only the forward part. Nonlinear coupling to the backward beam breaks the shape-preserving propagation in (III), since the angular momentum ($\alpha$) is not large enough to decouple the beams. The edges of the evanescent regime are marked by a black dashed rectangle in (IV), emphasizing in the enlarged window that the peak of the Fourier amplitude is at $k_x>k_0$. All figures have angular momentum of $\alpha=60$ and nonlinearity $\varepsilon_{mean}=0.05$, except for the linear plots which have $\alpha=45$.